\documentclass{emulateapj}
\usepackage{natbib}
\usepackage{graphicx}
\usepackage{amsmath}
\usepackage[version=3]{mhchem}
\usepackage{epsfig}
 \usepackage{color}
 \usepackage{times}

\begin{document}
%
\title{Formation and recondensation of complex organic molecules during
  protostellar luminosity outbursts}
\author{Vianney Taquet$^{1}$, Eva S. Wirstr\"om$^2$ and Steven B. Charnley$^{3}$}
%
\altaffiltext{1}{Leiden Observatory, Leiden University, P.O. Box 9513,
2300-RA Leiden, The Netherlands}
\altaffiltext{2}{Department of Earth and Space Sciences, Chalmers
  University of Technology, Onsala Space Observatory, SE-439 92
  Onsala, Sweden}
\altaffiltext{3}{Astrochemistry Laboratory and The
  Goddard Center for Astrobiology, Mailstop 691, NASA Goddard Space
  Flight Center, 8800 Greenbelt Road, Greenbelt, MD 20770, USA}
    \date{Received - ; accepted -}
\begin{abstract}

During the formation of stars, the accretion of the surrounding
material toward the central object is thought to undergo strong
luminosity outbursts, followed by long periods of relative
quiescence, even at the early stages of star formation when the
protostar is still embedded in a large envelope. 
We investigated the gas phase formation and the recondensation of
the complex organic molecules di-methyl ether and
methyl formate, induced by sudden ice evaporation processes occurring
during luminosity outbursts of different amplitudes in
protostellar envelopes. 
For this purpose, we updated a gas phase chemical network forming
complex organic molecules in which ammonia plays a key role. 
The model calculations presented here demonstrate that ion-molecule
reactions alone could account for the observed presence of di-methyl
ether and methyl formate in a large fraction of protostellar
cores, without recourse to grain-surface chemistry, { although they
depend on uncertain ice abundances and gas phase reaction branching
ratios.  }
In spite of the short outburst timescales of about one hundred years,
abundance ratios of the considered species with respect to methanol
higher than 10\% are predicted during outbursts due to their low binding
energies relative to water and methanol that delay their
recondensation during the cooling.  
 Although the current luminosity of most embedded protostars would be
too low to produce complex organics in hot core regions that can be
observable { with current sub-millimetric interferometers}, previous
luminosity outburst events would induce a formation of COMs in
extended regions of protostellar envelopes with sizes increasing by up
to one order of magnitude.    

 \end{abstract}



\section{Introduction}

Complex organic molecules (COMs) have been observed in high quantities
around protostars, in their so-called high-mass hot cores and low-mass
hot corinos \citep{Blake1987, Cazaux2003, Bisschop2007}. 
The presence of many COMs in the gas phase can be understood as due to
the evaporation of ices from dust grains. 
 In this case, atom addition reactions on grain surfaces could account
 for many of the organic molecules observed like
  CH$_3$OH, C$_2$H$_5$OH, CH$_3$CHO, or HCOOH \citep[e.g.][]{Herbst2009}. 
Other common organic molecules like di-methyl ether (DME) and methyl
formate (MF) appear to require either energetic processing of simple ices
containing methanol \citep{Oberg2009} or ion-molecule reactions
post-evaporation \citep{Charnley1995}.  
However, published astrochemical models tend to underestimate their
abundances with respect to methanol, their likely parent molecule
\citep{Taquet2015}.  
These models usually consider constant physical conditions,
representative of hot cores, or simple physical models of core
collapse inducing a gradual warm up of the protostellar envelope. 

However, { during the star formation process}, accretion of the surrounding
material toward the central protostar is thought to undergo
frequent and strong eruptive bursts inducing sudden increases of the
luminosity by one or two orders of magnitude, followed by long periods
of relative quiescence. 
As suggested by magnetohydrodynamics (MHD) simulations, luminosity outbursts could be due
to thermal, gravitational and magnetorotational instabilities
\citep{Bell1994, Zhu2009} or gravitational fragmentation in the circumstellar
disk \citep{Vorobyov2005, Vorobyov2015} and could explain the spread
in bolometric luminosities observed for low-mass embedded protostars
\citep[see][]{Dunham2012}.  
FUor and EXor objects, whose SEDs can be attributed to Class I/II
protostars but with heavier disks and higher accretion rates
\citep{Gramajo2014}, { probably} undergo such luminosity outbursts
\citep{Abraham2004}.  
The recent detection of an outburst toward the Class 0 protostar
HOPS383 by \citet{Safron2015} suggests that luminosity outbursts are also occurring
in the embedded Class 0 phase although they can be hardly observed
directly, due to their optically thick surrounding envelope. 
The strong and sudden increase of the temperature induced by the
luminosity outburst can significantly alter the chemical evolution in the
envelope and in the disk by triggering the fast evaporation of
solid species into the gas phase, resulting in an increase of their
gaseous abundances long after the system has returned into a quiescent stage.
The evaporation of icy species would therefore influence the abundances of
commonly observed molecules, such as N$_2$H$^+$ and HCO$^+$, whose
abundances are governed by CO or H$_2$O \citep{Visser2012}, an effect
proposed by \citet{Jorgensen2013} to explain the non-detection of
HCO$^+$ and the presence of CH$_3$OH toward the inner envelope of the
low-luminosity protostar IRAS15398-3359. 

The efficient ice evaporation induced by luminosity outbursts
could also trigger the gas phase formation of COMs. 
In addition, the low binding energy of some COMs with respect to water
and methanol ices would induce a differentiation in the recondensation,
altering the abundances of COMs with respect to these more simple species.
In this work, we investigate the hot core chemistry leading to the
formation of gaseous COMs and the impact of luminosity outbursts on
the formation and their subsequent recondensation of COMs by focusing
on the formation of the two { O-bearing} prototype COMs di-methyl
ether and methyl formate. We also compare our model predictions with
results from sub-millimetric observations toward low-mass to high-mass
protostars.  
{ Section 2 describes the model used in this work. Section 3
  presents the chemical evolution for constant physical conditions
  while Section 4 shows the chemical evolution during strong and weak
  luminosity outbursts. We discuss the implications of this work in
  Section 5 and outline the conclusions in Section 6.
}

\section{Model}

\subsection{Physical Model}

{
According to hydrodynamical models of disk instability and
fragmentation, the embedded Class 0 and Class I phases show a highly
variable evolution of their luminosity with various outbursts of
different amplitudes. 
In this work, we investigated how the chemical evolution is impacted
by two types of outburst whose properties are taken from the model
results by \citet{Vorobyov2015}:  
1) one strong outburst, increasing the luminosity by a factor of 100 from
2 to 200 $L_{\odot}$ every $\sim 0.5 - 1 \times 10^5$ yr;
2) a series of weak outbursts, increasing the luminosity by a factor
of 10 only from 2 to 20 $L_{\odot}$, but more frequently (every $\sim
5 \times 10^3$ yr).
For the two types of outburst, we assumed that the luminosity
instantaneously increases from $L_{\textrm{min}} = 2 L_{\odot}$ to its maximal
luminosity $L_{\textrm{max}}$ and then decreases exponentially
following the formula 
\begin{equation}
{L_{\star}(t) } = (L_{\textrm{max}}-L_{\textrm{min}}) \exp(-t/\tau) + L_{\textrm{min}}
 \end{equation}
 $\tau$ being the outburst timescale. The outburst duration, assumed
 to be the time during which the luminosity remains higher than half
 of its maximal luminosity, is highly variable. Models and
 observations show that it can vary between a  few decades to a few
 centuries depending on the type of predicted instabilities and
 observed source \citep[see][]{Audard2014}. 
We therefore varied $\tau$ between 75 and 300 yr, $\tau = 150$ yr,
corresponding to an outburst duration of $\sim 100$ yr, being our
standard value.    

Figure \ref{nHtemprad} presents the dust temperature structure in the
envelope surrounding Serpens-SMM4, a typical Class 0 protostar with a
current bolometric luminosity $L_{\textrm{bol}} = 2 L_{\odot}$ and an envelope
mass of 2.1 $M_{\odot}$ before and during the two types of outburst
considered in this work, as computed by the radiative transfer code
DUSTY \citep{Ivezic1997}. The protostar properties were derived by
\citet{Kristensen2012} and we followed the methodology described in
\citet{Taquet2014} to compute the temperature structure. 
\citet{Johnstone2013} found that the time needed by the dust to heat
up in response to a luminosity outburst is much shorter, typically a
few days to a few months, than the typical duration of a luminosity
outburst. We therefore assumed that the dust temperature
instantaneously scales with the luminosity evolution. 
From Fig. \ref{nHtemprad}, it can be seen that the dust temperature
roughly follows the Stefan-Boltzmann's law throughout the envelope
\begin{equation}
T(t) = T_{\textrm{min}} \times (L_{\star}(t)/L_{\textrm{bol}})^{1/4}
\end{equation}
where $T_{\textrm{min}}$ is the temperature before and at the
end of the outburst. 
The two types of outburst expand the hot-core region, where ices are
thermally evaporated at $T = 100 - 120$ K, from 15 AU to 60 AU for
weak outbursts and to 150 AU for strong outbursts. 
We therefore considered several pre-outburst temperatures of
particular interest:
$T_{\textrm{pre}} = 100$ K, which is slightly lower than the evaporation/condensation
temperature of methanol; 
$T_{\textrm{pre}} = 70$ K which is slightly lower than the evaporation/condensation
temperature of DME and MF;
and $T_{\textrm{pre}} = 40$ K, giving $T_{\textrm{max}} = 125$ K which is
sufficient to thermally evaporate the whole content of ices.
Three values of density were chosen, in order to represent the
different densities expected to be found in the inner regions of
Class 0 and Class I protostellar envelopes where the temperature is
between 40 and 100 K \citep{Kristensen2012}:
$n_{\textrm{H}} = 5 \times 10^6, 5 \times 10^7, 5 \times 10^8$
cm$^{-3}$.  
We also adopted three values for the size of interstellar grains: 
$a_{\textrm{d}} = 0.05$ $\mu$m, representing the grain size needed to
match the integrated surface area observed in the diffuse ISM;
$a_{\textrm{d}} = 0.2$ $\mu$m, the upper limit of the grain size
distribution observed in the diffuse ISM { and commonly used by
astrochemical models};
and $a_{\textrm{d}} = 1$ $\mu$m, a higher grain size obtained through grain
growth in dense cores as observed by \citet{Pagani2010}. 

\begin{figure}[htp]
\centering 
\includegraphics[width=0.49\textwidth]{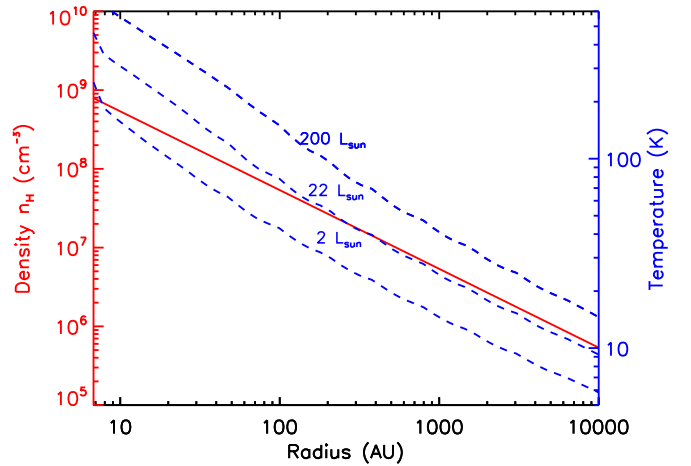} 
\caption{Dust temperature structure in the envelope surrounding the
  Class 0 protostar Serpens-SMM4, with a bolometric luminosity
  $L_{\textrm{bol}} = 2 L_{\odot}$ and an envelope mass of 2.1
  $M_{\odot}$  before and during the two types of luminosity outburst
  considered in this work.   }
\label{nHtemprad}
\end{figure}
}

\subsection{Chemical Model}

\begin{figure*}[htp]
\centering 
\includegraphics[width=0.8\textwidth]{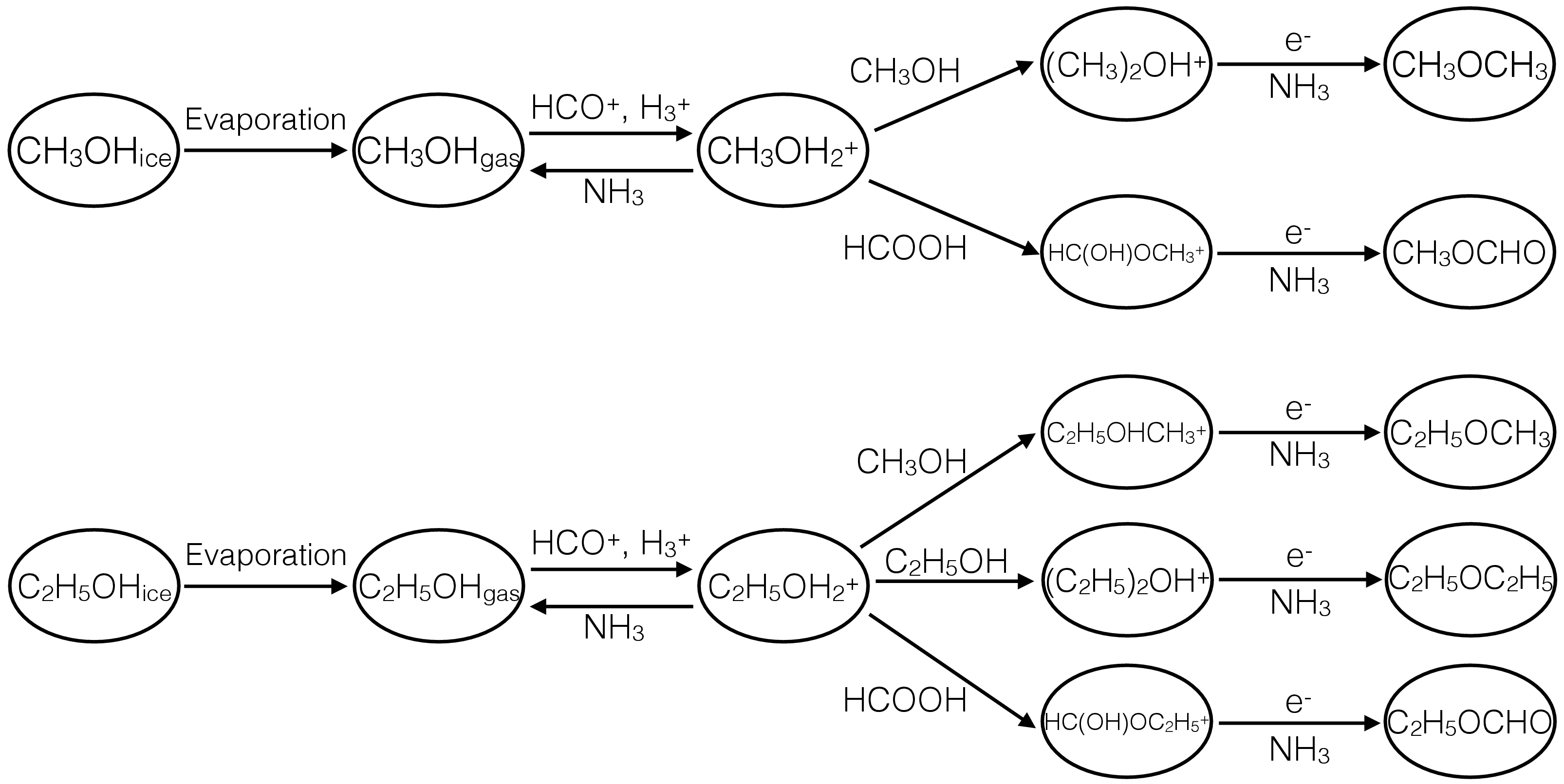}
\caption{Schematic picture of the gas phase chemical network used in
  this work to produce the complex organic molecules dimethyl ether,
  methyl formate, methyl ethyl ether, diethyl ether and ethyl formate
  from the evaporation of methanol and ethanol. }
\label{chemnet}
\end{figure*}

The chemistry is followed as a function of time with the GRAINOBLE
gas-grain astrochemical model presented in previous studies
\citep{Taquet2012, Taquet2014}.  
In this work, we focus our study on the gas phase chemistry and the
gas-grain interactions through freeze-out and thermal
evaporation. 
We used an updated version of the chemical network described in
\citet{Rodgers2001}. 
The rate of several key reactions has been updated while new reactions
have been added following recent experimental and theoretical works. 
{ The formation of complex organics through ion-neutral gas phase
  chemistry is triggered by the protonation of evaporated ices, and of
  methanol in particular. 

\begin{table}[htp]
\centering
\caption{Initial abundance, binding energy, and proton affinity of
  selected species. } 
\begin{tabular}{l c c c c c c}
\hline
\hline
Species & $n_{\textrm{ini}}/n_{\textrm{H}}$ & $E_{b,\textrm{bare}}$ & $E_{b,\textrm{wat}}$& $E_{b,\textrm{pure}}$& Ref. ($E_b$) & PA$^1$ \\
 & & (K) & (K) & (K) & & (kJ/mol) \\
\hline
H$_2$O & $1 \times 10^{-4}$ & 1870 & 5775 & 5775 & 2, 3 & 689 \\
CO & $3.8 \times 10^{-5}$ & 830 & 1150 & 855 & 4, 5, 6 & 593 \\
N$_2$ & $1.6 \times 10^{-5}$ & 790 & 790 & 790 & 7 & 494 \\
CO$_2$ & $3.0 \times 10^{-5}$ & 2270 & 2690 & 2270 & 5, 8 & 539 \\
CH$_4$ & $5.0 \times 10^{-6}$ & 1090  &  1090 & 1090 & 9 & 544 \\
NH$_3$ & $5.0 \times 10^{-6}$ & 5535 & 5535 & 3075 & 4, 10 & 854 \\
H$_2$CO & $2.5 \times 10^{-6}$ & 3260 & 3260 & 3765 & 11 & 713 \\
CH$_3$OH & $7.0 \times 10^{-6}$ & 5530 & 5530 & 4930 & 4, 12 & 754 \\
HCOOH & $1.6 \times 10^{-6}$ & 5570 & 5570 & 5000 & 4, 13 & 743 \\
C$_2$H$_5$OH & $1.6 \times 10^{-6}$ & 6795 & 6795 & 5200 & 14, 13 & 776 \\
CH$_3$OCH$_3$ & 0 & 4230 & 4230 & 3300 & 14, 13 & 792 \\
CH$_3$OCHO & 0 & 4630 & 4630 & 4000 & 14, 13 & 782 \\
C$_2$H$_5$OCHO & 0 & 5895 & 5895 & 4900 & 15 & 799 \\
CH$_3$OC$_2$H$_5$ & 0 & 5495 & 5495 & 4400 & 16 & 809 \\
C$_2$H$_5$OC$_2$H$_5$ & 0 & 6760 & 6760 & 5100 & 17 & 828 \\
CH$_3$CN & 0 & 4680 & 4680 & 4680 & 14, 13 & 781 \\
\hline
\end{tabular}
\label{abu_ini}
\tablecomments{
$^1$: The proton affinities (PA) are taken from the NIST Chemistry
WebBook (http://webbook.nist.gov/chemistry/)
$^{2}$: \citet{Avgul1970}; 
$^{3}$: \citet{Fraser2001};
$^4$: \citet{Collings2004};
$^5$: \citet{Noble2012a};
$^6$: \citet{Acharyya2007};
$^7$: \citet{Bisschop2006};
$^{8}$: \citet{Sandford1990}; 
$^{9}$: \citet{Herrero2010}; 
$^{10}$: \citet{Sandford1993}; 
$^{11}$: \citet{Noble2012b};
$^{12}$: \citet{Brown2009};
$^{13}$: \citet{Oberg2009}; 
$^{14}$: \citet{Lattelais2011}; 
$^{15}$: $E_b$ = $E_b$(CH$_3$OCHO) + $E_b$(C$_2$H$_5$OH) - $E_b$(CH$_3$OH);
$^{16}$: $E_b$ = $E_b$(CH$_3$OCHO$_3$) + $E_b$(C$_2$H$_5$OH) - $E_b$(CH$_3$OH);
$^{17}$: $E_b$ = $E_b$(CH$_3$OCHO$_3$) + $2 \times$ ($E_b$(C$_2$H$_5$OH) - $E_b$(CH$_3$OH))
} 
\end{table}

Electronic recombination (ER) reactions involving the protonated ions
associated with methanol, formic acid, DME, and ethanol
have been measured by \citet{Geppert2006, Hamberg2010a, Hamberg2010b,
  Vigren2010}. For the four systems, the total rate of the reaction
follows the expression 
$k(T) \sim 10^{-6} (T/300)^{-0.7}$ cm$^{-3}$ s$^{-1}$ while the
recombination leading to the complex organic molecule in consideration
has a low branching ratio between 6 and 13 \%, most of the reactions
being dissociative. 
To our knowledge, no experimental study focusing on the ER of protonated
MF has been published so far.  
We assumed the same rate and branching ratio as for the ER of
protonated DME measured by \citet{Hamberg2010a}.  

As in \citet{Rodgers2001}, we included proton transfer (PT) reactions
involving major ice species and abundant complex organics listed in
Table \ref{abu_ini} if they are thought to be exothermic. 
The exothermicity of the PT reaction
\begin{equation}
\ce{AH+} + \ce{B} \rightarrow \ce{A} + \ce{BH+}
\end{equation}
is given by the difference of proton affinities (PA) of B and A. The
reaction will therefore occur if the proton  
affinity of B is higher than that of A. Table \ref{abu_ini} lists the
proton affinity of the major ice species and relevant complex
organics. 
In particular, ammonia NH$_3$ easily reacts with most protonated ions through
exothermic barrierless PT reactions due to its high proton affinity. 
{ PT reactions can be in competition with charge transfer and
  condensation reactions \citep[see][]{Huntress1977} while
  dissociative proton transfer can occur when the PA of the acceptor
  greatly exceeds that of the donor, but the dissociation can only on
  the acceptor molecule \citep[see][]{Smith1994}.}
{ \citet{Hemsworth1974} experimentally studied at 297 K the
  reactivity of 11 reactions between ammonia and the protonated
  counterpart of neutral molecules with lower PA and of different
  complexity, from   H$_2$ to C$_4$H$_8$. 
They showed that all the studied reactions led to non-dissociative PT
reactions, the formation of NH$_4^+$ appearing to be the 
dominant ($\geq 90$ \%) channel in each case, which occur at the
collisional rate of $\sim 2 \pm 1 \times 10^{-9}$ cm$^3$ s$^{-1}$.
In a latter study, \citet{Feng1994} experimentally studied reactions
involving protonated formic acid and 11 neutral complex organic
molecules of higher proton affinities, such as methanol, methyl
cyanide, acetone and even more complex species like dimethyoxyethane
CH$_3$OCH$_2$CH$_2$OCH$_3$. All the reactions were also found to be
non-dissociative ($\geq 99$ \%) proton-transfer reactions occurring at
the collisional rate $\sim 2 \pm 1 \times 10^{-9}$ cm$^3$ s$^{-1}$
\citep[see][for a more exhaustive list of experimentally studied PT
reactions]{Anicich1993}. 
}
{ Following these experimental results}, we assumed that all the PT reactions
introduced in the chemical network are non-dissociative and occur with
a rate of $2 \times 10^{-9}$ cm$^3$ s$^{-1}$.   

Experiments and quantum calculations show that that the reaction
between CH$_3$OH$_2^+$ and H$_2$CO does not lead to protonated methyl
formate \citep{Karpas1989, Horn2004}.
In this model, protonated methyl formate is instead formed through the
barrierless methyl cation transfer 
reaction  \citep{Ehrenfreund2000} 
\begin{equation} 
\ce{CH_3OH2+ + HCOOH} \rightarrow \ce{HC(OH)OCH_3+ +  H_2O}. 
\label{MFreac}
\end{equation}
Experimental and theoretical studies of this reaction system indicate
that the {\it trans} conformer of protonated methyl formate should be
produced and that the channel forming the more excited {\it cis}
conformer of protonated MF has an activation barrier of 1320 K
\citep{Neill2011, Cole2012}.  
The rate of this reaction has been measured experimentally by
\citet{Cole2012} who measured a predominant branching
ratio of 95 \% for adduct ion products and a low branching ratio of 5
\% for the reaction leading to protonated MF. 
However, \citet{Cole2012} suggested that the branching ratio for the
adduct ion would be lowered in the ISM due to the lower pressures
found in the ISM with respect to the pressure obtained at the lab. We therefore
assumed a branching ratio of 100 \% for the reaction leading to
$\ce{HC(OH)OCH_3+}$.
According to the proton affinities of ({\it cis}-)MF and \ce{NH3}, the
PT reaction between {\it cis}-protonated MF and \ce{NH3}  is
barrierless and has a high exothermicity of $\sim 70$ kJ/mol (see
Table \ref{abu_ini}).  
The energy difference between the {\it cis}- and {\it
  trans}-protonated MF conformers is only about 25 kJ/mol
\citep{Neill2011}, the reaction between {\it trans}-protonated MF and
\ce{NH3}, forming either {\it cis}-MF or {\it trans}-MF, is therefore
also likely exothermic.  
Since there are, to our knowledge, no quantitative data on the
branching ratios for the formation of {\it cis}- or {\it trans}-MF via
this reaction, we assumed a branching ratio of 100 \% for the
formation of the more stable {\it cis-}MF conformer, which is more
stable than the {\it trans-} conformer by 25 kJ/mol. }
{ The gas phase chemistry forming other O-bearing complex
  organic molecules from the evaporation of methanol and ethanol is
  summarized in Figure \ref{chemnet}, and follows the experimental
  results of \citet{Karpas1989} \citep[see also][]{Charnley1995}. 
Dimethyl ether, methyl ethyl ether, and diethyl ether are formed from
reactions between protonated methanol or protonated ethanol with
methanol or ethanol. We assumed that the reaction between protonated
ethanol and formic acid, forming protonated ethyl formate, has the
same rate than the reaction between methanol and formic acid.}
In total, the chemical network consists in 325 species and 2787
chemical processes. 

\subsection{Initial Abundances and Binding Energies}

{ 
For each species $i$, the effective binding energy $E_b(i)$ relative to
the surface is given by the additive contribution of the binding energy
relative to a bare grain substrate, an Amorphous Solid Water (ASW),
and a pure ice $i$ according to their fractional coverage in the ice,
following the methodology described in \citet{Taquet2014}.
The binding energies of the main ice components and some abundant COMs
with respect to the three substrates have been measured in laboratory
experiments and are listed in Table \ref{abu_ini}.  
Differences in the binding energies of complex organics can be
noticed, leading to different temperatures of sublimation and
recondensation. 
For example, as MF and DME show lower binding energies than methanol,
they will evaporate and recondense at lower temperatures of 70-80 K
with respect to methanol, but also to formic acid or water, which
recondense at 100-110 K. 
In contrast, ethanol C$_2$H$_5$OH has a higher binding energy and evaporates
at a higher temperature of 120 K. 
}

The initial abundances of molecular ices are taken from infrared
observations of ices and are listed in Table \ref{abu_ini}. The water
abundance of $10^{-4}$ with respect to H nuclei follows ice
observations by \citet{Tielens1991} and
\citet{Pontoppidan2004}. Abundances of solid CO, CO$_2$, CH$_4$,
NH$_3$, and CH$_3$OH are taken from the abundance medians derived by
\citet{Oberg2011} toward a sample of protostars. 
{ Theoretical models suggest that H$_2$CO, HCOOH, and C$_2$H$_5$OH
  should also be present in cold interstellar ices \citep[see][]{Herbst2009}.
However, their abundances in ices  are highly}
uncertain since the features used for their detection are contaminated
by other mixtures. We fixed the H$_2$CO abundance to 2.5 \%, following the
abundance of the C1 component attributed to H$_2$CO+HCOOH. 
The HCOOH abundance has been derived from the band feature at 7.25
$\mu$m detected by \citet{Boogert2008} toward 12 low-mass protostars,
giving a mean abundance of 3.2 \% with respect to water. However,
C$_2$H$_5$OH appears to be a plausible carrier for this feature as well
\citep[see][]{Oberg2011}. We therefore assumed an abundance of 1.6 \%
for HCOOH and C$_2$H$_5$OH.

\section{Chemistry during Constant physical conditions}

\subsection{Impact of proton transfer reactions}

\begin{figure*}[htp]
\centering 
\includegraphics[width=0.8\textwidth]{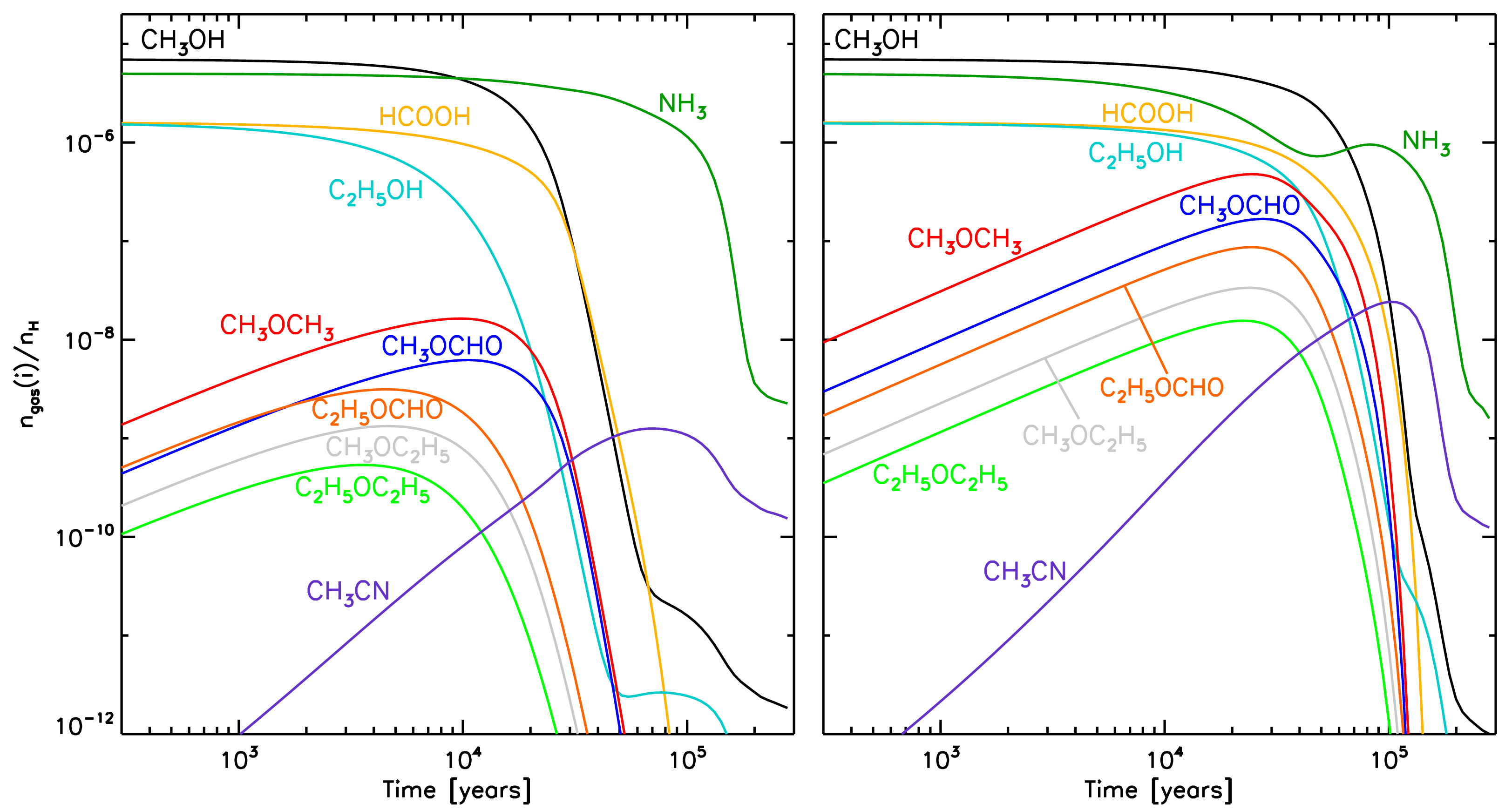}
\caption{Temporal evolution of the absolute abundances of complex
  organics by   neglecting (left) and including (right) the proton
  transfer reactions with ammonia for $n_{\textrm{H}} = 5 \times 10^7$
  cm$^{-3}$, $T = 150$ K,   $\zeta = 3 \times 10^{-17}$ s$^{-1}$, $A_{\textrm{V}} = 20$ mag. }
\label{X_constant}
\end{figure*}

{ We show in Figure \ref{X_constant} the impact of the proton transfer
reactions involving NH$_3$ introduced in the chemical network on the
formation and destruction of COMs for constant physical conditions: 
$n_{\textrm{H}} = 5 \times 10^7$ cm$^{-3}$, $T = 150$ K, $\zeta = 3
\times 10^{-17}$ s$^{-1}$, $A_{\textrm{V}} = 20$ mag, assumed to be our
standard physical parameters.}

After their evaporation into the gas phase, ice species, such as
CH$_3$OH, HCOOH, C$_2$H$_5$OH, are protonated through proton-transfer
reactions involving the abundant ions H$_3$O$^+$ or HCO$^+$. 
CH$_3$OH$_2^+$ can then react with CH$_3$OH, or HCOOH via methyl
cation transfer reactions to form the DME and MF protonated ions.   
{ Since the ER reactions involving the protonated COM ions lead
  predominantly to their break-up into small molecules, complex
  organics are not formed efficiently through gas phase chemistry
  if PT reactions involving NH$_3$ are neglected, their abundances not
  exceeding $10^{-8}$. Moreover, they are quickly destroyed in less than a few
  $10^4$ yr into small fragments. 
The incorporation of the PT reactions involving NH$_3$ increases the COM
abundances by one to two orders of magnitude. PT reactions, which are
likely non-dissociative, also delay the destruction of COMs since they
dominate over the dissociative ER reactions.}
DME and MF reach an abundance peak of $5 \times 10^{-7}$ and $2 \times
10^{-7}$ in $2-3 \times 10^4$ yr, respectively { and start to be
  efficiently destroyed after $5 \times 10^4$ yr. }
MF reaches a slightly lower abundance than DME due to the lower
abundance of HCOOH relative to CH$_3$OH. 
%
{ A similar chemistry triggered by the protonation of
  ethanol produces methyl ethyl ether, di-ethyl ether and ethyl formate
as depicted in Fig. \ref{chemnet} \citep[see also][]{Charnley1995}.  
As seen in Table \ref{abu_ini} ethanol has a higher proton affinity
than methanol. The PT reaction between protonated methanol and neutral
ethanol therefore enhances the ethanol protonation with respect to
methanol and induces a more efficient conversion from ethanol to the
longer COMs, like ethyl formate, methyl ethyl ether or di-ethyl
ether. Their abundance remains nevertheless lower than MF and 
DME because of the lower initial abundance of ethanol. } 

{ The incorporation of the new PT reactions also tends to enhance
  the destruction of NH$_3$ as protonated methanol becomes the main
  proton donor of NH$_3$. However, NH$_3$ survives for a longer time
  than other more complex species because the ER reaction involving
  NH$_4^+$ is mostly non-dissociative and reforms either NH$_3$ or
  NH$_2$. Moreover, NH$_2$ can also be protonated to form NH$_3^+$
  that reforms   NH$_3$ through the reaction between NH$_3^+$ and H$_2$. 
  Since this latter process is in competition with the reaction
  between NH$_2$ and H, whose rate increases with the temperature,
  higher temperatures tend to increase the destruction efficiency of ammonia.} 
Our gas phase chemical network also produces methyl cyanide CH$_3$CN,
from the reaction between HCN and CH$_3^+$ through protonated methyl
cyanide but in lower abundances ($\sim 10^{-8}$) and obtained in a longer
time ($\sim 10^5$ yr).

\subsection{Impact of other parameters}

\begin{figure*}[htp]
\centering 
\includegraphics[height=98mm]{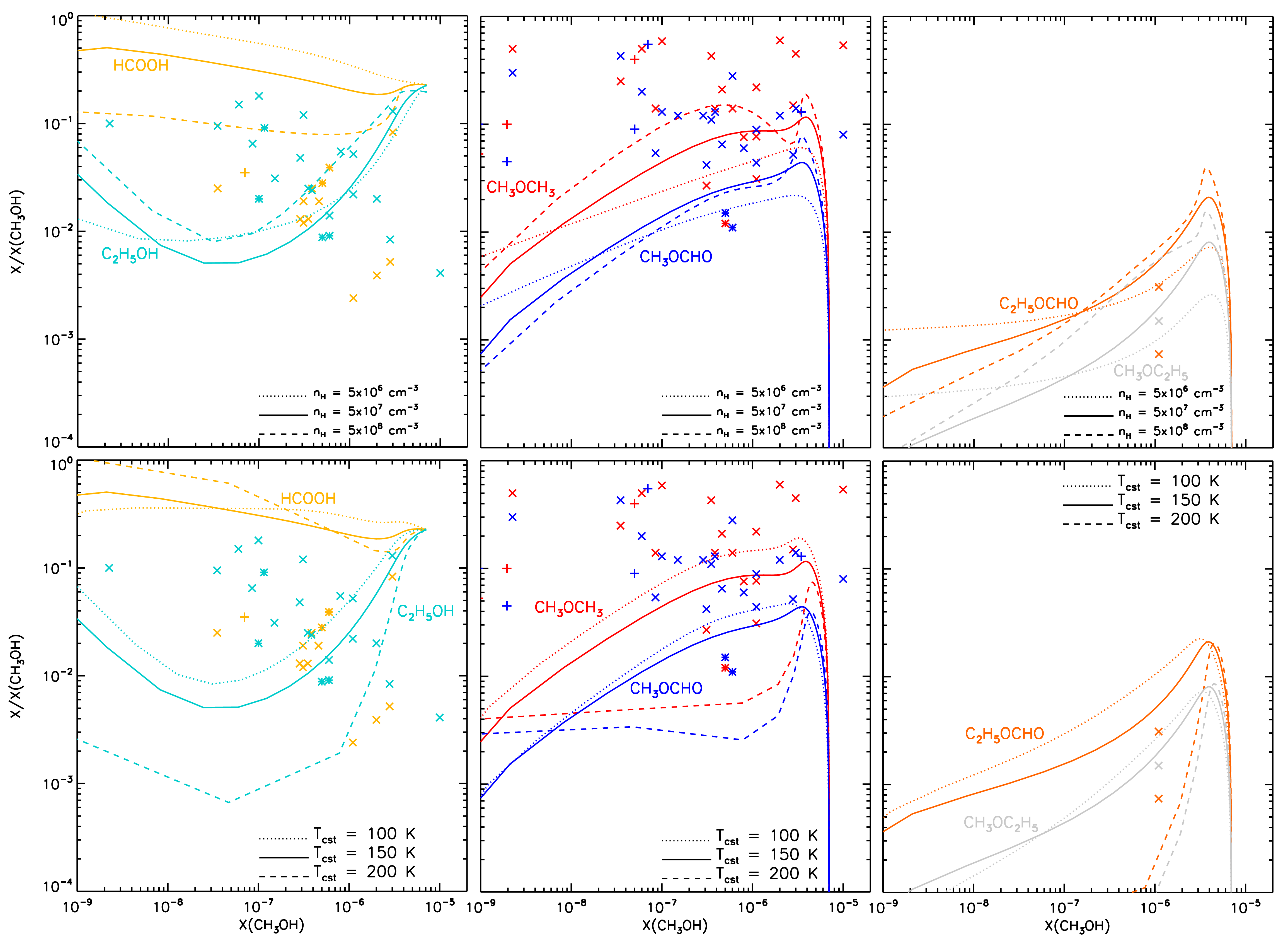} 
\caption{Evolution of the gaseous abundances of formic acid, ethanol (left panels),
  di-methyl ether, methyl formate (center panels), ethyl formate and
  methyl ethyl ether (right panels) relative to methanol with the absolute
  CH$_3$OH abundance assuming constant physical conditions. 
Top and bottom panels show the effect of the density
$n_{\textrm{H}}$, and the temperature $T_{\textrm{cst}}$, respectively, on the chemistry. 
Pluses, stars, and crosses represent the ratios observed toward
low-mass, intermediate-mass, and high-mass protostars, respectively,
summarised in \citet{Taquet2015}.}
\label{Xmeth_constant}
\end{figure*}

In order to compare our model predictions to observations
of complex organics toward protostars, we show in Figure
\ref{Xmeth_constant} the evolution of the predicted abundances of
{ formic acid, ethanol, DME, MF, ethyl formate, and methyl ethyl-ether}
with respect to methanol as a function of the absolute methanol 
abundance { for different values of the total density, and temperature.}

With time, the abundance ratios evolve from the right to the
left of each panel, { their exact evolution depending on the
  initial abundance in ices and their chemistry in the gas phase}. 
{ Formic acid and ethanol both have an initial abundance of $\sim
  20$ \% with respect to methanol according to the standard assumed
  abundances (see Table \ref{abu_ini}). The temporal evolution of their
  abundance, however, shows opposite trends because of their different
  proton affinities. 
{ Ethanol has a higher proton affinity than
  methanol, allowing the proton-transfer reaction between protonated
  methanol and ethanol to occur. The protonation of ethanol and its
  conversion to   larger species will therefore be enhanced with
  respect to methanol.
Its abundance ratio therefore decreases to $\sim 1$ \% for
  the standard model (solid lines in Fig. \ref{Xmeth_constant}). 
On the other hand, formic acid has a lower proton affinity than
methanol, methanol will consequently limit its protonation through the
proton transfer reaction between protonated formic acid
methanol and therefore its destruction to larger species. Its
abundance remains constant for a longer time than methanol, its
abundance ratio slightly increasing from $\sim 20$ to $\sim 50$ \%.}}
{ Gas phase chemistry produces high abundance ratios of MF and DME, 
  reaching peaks of 10 and 4 \% respectively at $2-3 \times   10^4$ yr
  when the methanol abundance is still high ($> 10^{-6}$). 
The dissociative ER reactions involving the protonated COMs have
slightly higher rates than the ER reactions destroying protonated
methanol ($1.8 \times 10^{-6}$ for DME vs $0.9 \times 10^{-6}$ cm$^{-3}$
s$^{-1}$ for methanol at 300 K), the MF and DME abundance ratios
therefore slowly decrease with the destruction of methanol and other
large molecules at longer timescales. }   
{ Abundance ratios of ethyl formate and methyl ethyl ether show
  similar trends than MF and DME but are lowered by one order
  of magnitude, due to the lower initial abundance of ethanol.}

{ 
The abundance ratios also depend on the assumed physical conditions.
The density slightly influences the efficiency of the COMs formation
since an increase of density from $5 \times 10^6$ to $5 \times 10^8$
cm$^{-3}$ slightly increases the MF and DME abundances, but only by a
factor of 4 at maximum, due to the decrease of the abundance of
electrons and protonated ions that destroy neutral COMs.  
The abundance of COMs does not depend on the
temperature for values lower than 150 K, their abundances only varying
by a factor of 2 at maximum. However, higher temperatures enhance the
destruction of COMs, decreasing the MF and DME abundance ratios by one
order of magnitude between 150 and 200 K due to the more
efficient destruction of NH$_3$, as explained in section 3.1.
}

\begin{figure}[htp]
\centering 
\includegraphics[width=85mm]{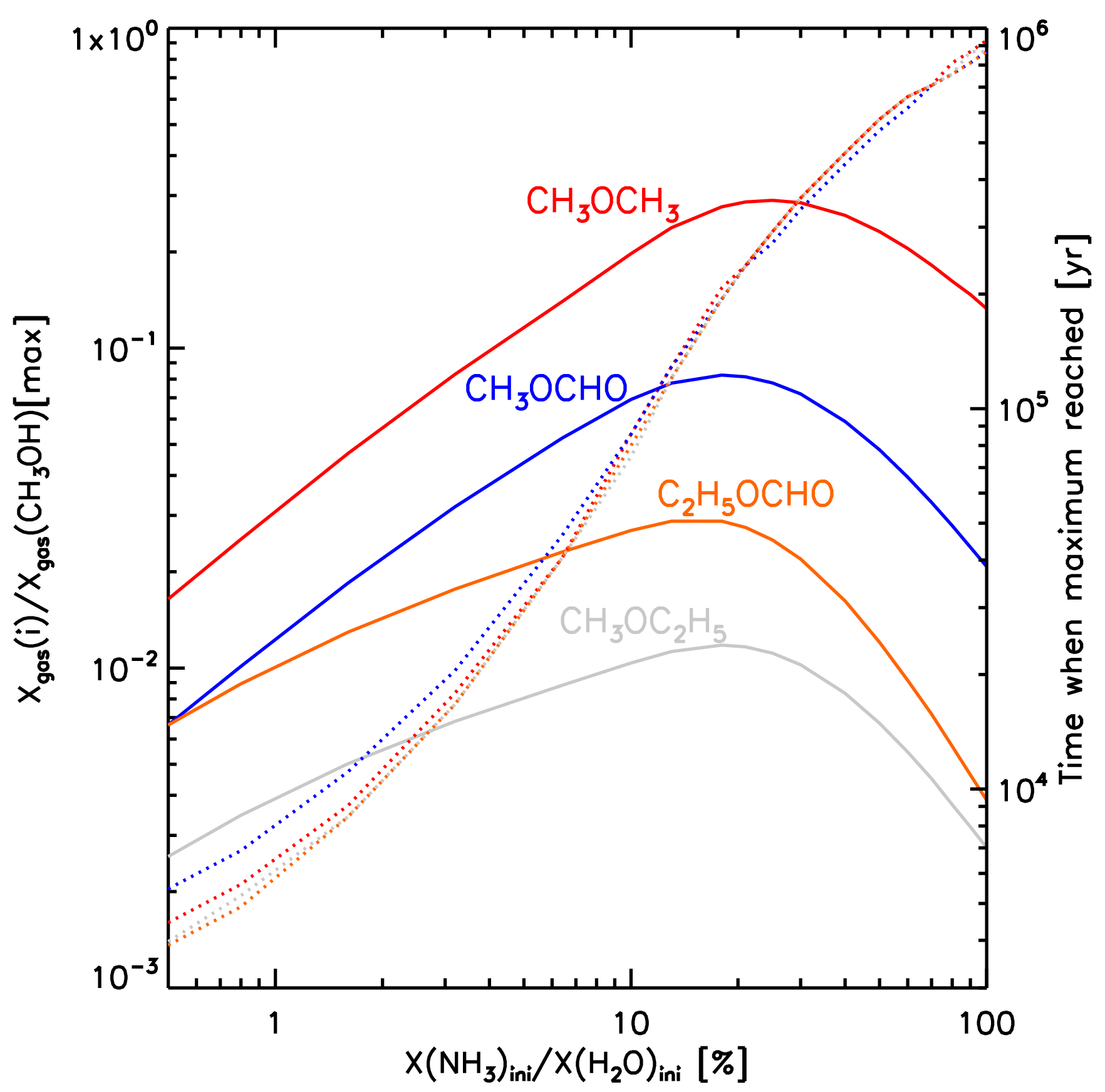} 
\caption{Maximum di-methyl ether, methyl formate, ethyl formate, and
  methyl ethyl ether abundances relative to methanol (solid lines) and
time when the maximum is reached (dotted lines) as a function of the initial
abundance of ammonia.}
\label{Xch3oh_Xnh3}
\end{figure}

{ However, } the most important parameter is the initial abundance
of ammonia { injected in the gas phase} as it governs the
efficiency of proton exchange reactions both with CH$_3$OH$_2^+$ and
with protonated COM ions.  
{ Figure \ref{Xch3oh_Xnh3} shows the maximal abundance relative to
methanol reached by DME, MF, ethyl formate, and methyl ethyl ether and
the time required to reach their maximal abundances as function of the
initial abundance of ammonia.}
On the one hand, a low NH$_3$ abundance induces an efficient protonation of
methanol but also a low formation of COMs from large ions, possibly
mostly by ERs.
On the other hand, a very high NH$_3$ abundance keeps the protonated
methanol at such low levels that both the destruction of methanol by
dissociative electron recombination, and the formation of COMs, are
limited \citep{Rodgers2001}.
Consequently, abundances of COMs increase with ammonia
abundances between X(NH$_3$)/X(H$_2$O) = 0 \% and 20 \% with respect
to water to reach peaks of 30 and 8 \% { for DME and MF}, respectively and
then decline for higher ammonia abundances. 
{ However, the time needed to reach the maximal abundances strongly
  increases with the initial ammonia abundance because it delays the
  protonation of methanol. }
{ A reasonable amount of solid NH$_3$, with similar abundances to
  those measured in interstellar ices toward low-mass protostars
  ($X$(NH$_3$)/$X$(H$_2$O) = 5 - 10 \%), provides the best balance
  between efficient   methanol protonation and efficient formation of
  neutral COMs from protonated ions, enhancing the gas phase
  production of O-bearing COMs.  
On the other hand, a high initial abundance of solid ammonia of 25 \%,
as assumed by \citet{Rodgers2001}, inhibits the conversion of
methanol and ethanol to more complex species as it requires too much
time \citep[$4 \times 10^5$ yr compared to the Class 0 lifetime of $\sim
10^5$ yr][]{Evans2009}.}

{
Infrared observations of interstellar ices suggest that solid methanol
shows a strong variation of its abundance, from less than 3 \% in a
significant number of low-mass protostars to more than 30 \% in a few
massive sources \citep{Gibb2004, Bottinelli2010, Oberg2011}. Although
the origin of the 7.25 $\mu$m band is still a matter of debate, HCOOH
abundances derived from observations of this band toward low- and
high-mass protostars is also highly variable, between less than 0.5 \%
to more than 7 \%. 
The gas phase abundance of the COMs studied in this work obviously
depends on the initial abundance of solid species injected in the gas
phase.
The maximal abundance ratios reached by the daughter COMs depend on
the ratio between the initial methanol (or ethanol) abundance relative
to ammonia. A low methanol abundance relative to NH$_3$ limits its
protonation due to the high NH$_3$ abundance that reforms back
methanol while gas phase chemistry is not efficient enough to produce
a high abundant of COMs when the methanol abundance is higher. 
It is found that DME and MF abundance ratios reach their maximum when
methanol and ammonia have a similar abundance of 5-10 \%. 
}

\section{Chemical evolution during luminosity outbursts}

\subsection{Strong luminosity outbursts}

{ This section describes the chemical evolution induced by one strong
  luminosity outburst, in which the central luminosity increases from
  2 to $200$ $L_{\odot}$, inducing an increase of the temperature by a
  factor of $\sim 3.2$, as explained in section 2.1. 
  The chemical evolution occurring during
  luminosity outbursts strongly depends on the binding energy of
  neutral species. We will therefore focus our study on di-methyl ether and
  methyl formate whose binding energies have been comprehensively
  studied experimentally by different groups, in contrast to the heavier
  species ethyl formate, methyl ethyl ether or di-ethyl ether {
    whose binding energies are guessed values}.}
The sudden increase in temperature induced by the luminosity
outburst triggers the evaporation of all icy species into the gas
phase, allowing an efficient formation of daughter COMs, such as DME
and MF, through the gas phase chemistry described in the previous
section. 

\begin{figure*}[htp]
\centering 
\includegraphics[height=98mm]{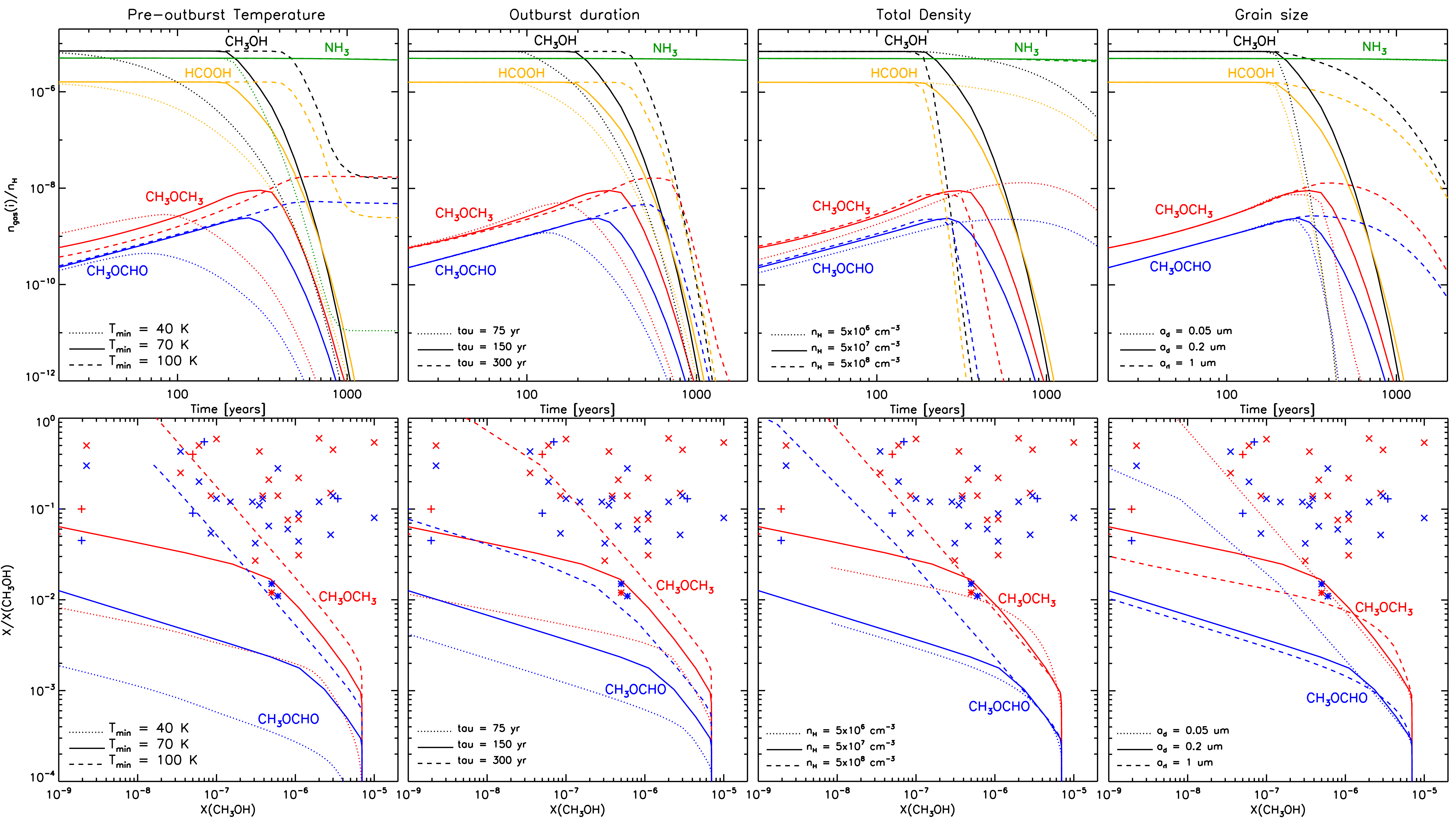}
\caption{
Evolution of the absolute abundances of methanol, ammonia, formic
acid, di-methyl ether and methyl formate with time (top panels) and of the
\ce{CH_3OCH_3}/\ce{CH_3OH} (red) and \ce{CH_3OCHO}/\ce{CH_3OH} (blue)
abundance ratios with the absolute CH$_3$OH abundance (bottom panels)
during one strong luminosity outburst. 
Pluses, stars, and crosses represent the ratios observed toward
low-mass, intermediate-mass, and high-mass protostars, respectively,
summarised in \citet{Taquet2015}.
Left, middle-left, middle-right and right panels show the influence of 
the pre-outburst temperature $T_{\textrm{min}}$, the luminosity
outburst timescale $\tau$, the total density $n_{\textrm{H}}$, and the
grain size $a_{\textrm{d}}$ respectively, on the chemistry. 
}
\label{X_outbursts}
\end{figure*}

Figure \ref{X_outbursts} presents the temporal evolution
of abundances of complex organics during one strong luminosity
outburst. 
{ In all panels, the solid curves show the chemical evolution
  during the model using the standard parameters ($n_{\textrm{H}} = 5
  \times 10^7$ cm$^{-3}$, $T_{\textrm{min}} = 70$ K, $\tau = 150$ yr,
    $a_{\textrm{d}} = 0.2$ $\mu$m). The dashed and dotted curves show
    the chemical evolution where one parameter is varied at a time.}
The formation of COMs is { efficient but limited by the short
  timescale of the outburst and the fast decrease of the temperature
  that induces a rapid recondensation of neutral species, and of
  methanol in particular
Absolute abundances reached during outbursts are consequently lower
than those obtained for constant physical conditions.} 
DME and MF reach abundance peaks of $\sim 10^{-8}$ only, the exact
value of the maximal abundance depending on the assumed physical
parameters.     
The absolute abundances of COMs tend to increase with the pre-outburst
temperature and the outburst timescale, since they { directly
  affect} the time spent by ices in the gas phase before their
recondensation during the cooling.   
The increase of the poorly constrained luminosity outburst timescale
from 75 to 300 yr or the pre-outburst temperature from 40 to 100 K
increases the DME and MF abundancess by one order of magnitude {
  simply due to the delay of the recondensation of neutral species.}

As shown in Table \ref{abu_ini}, MF and DME have lower binding
energies than methanol by $\sim 1000$ K, inducing a difference of
20-30 K in their temperature of recondensation. 
{ Methanol starts to condense at $100-110$ K whereas MF and DME
  freeze-out at a lower temperature of $70-80$ K.} 
The impact of the binding energy differences is illustrated in the
bottom panels of Fig. \ref{X_outbursts}, showing the evolution of the
DME and MF abundances with respect to methanol as function of the 
methanol abundance.
The formation efficiency of COMs is limited by the short luminosity
outburst timescales, inducing low abundance ratios at high methanol
abundances. However, the low binding energy of COMs delays their
freeze-out with respect to methanol. 
{ If the outburst timescale is longer than the freeze-out
  timescale, given by this formula
\begin{equation}
\tau_{\textrm{fr}} = 100 \textrm{ yr } \frac{5 \times 10^7
  \textrm{ cm}^{-3}}{n_{\textrm{H}}} 
 \frac{a_{\textrm{d}}}{0.2 {\textrm{ $\mu$m}}},
\label{freezeout}
\end{equation}
then the methanol abundance efficiently decreases during the cooling
before the onset of the recondensation of the more volatile DME and
MF, increasing their abundance ratio.
As a consequence, the evolution of the abundance ratio of COMs also
strongly depends on the total density of H nuclei and the grain size.
Higher densities or smaller grain sizes increase the freeze-out rate
of neutral species like methanol. This induces a slight decrease of absolute
abundances of COMs because methanol spends slightly less time in the
gas phase but, more importantly, a strong increase of their abundance
ratios by more than one order of magnitude as soon as the freeze-out
timescale becomes shorter than the outburst timescale.
According to equation (\ref{freezeout}), a high density of $5 \times
10^{8}$ cm$^{-3}$ or a small grain size of 0.05 $\mu$m decrease the
freeze-out timescale to less than a few decades, inducing an efficient
depletion of methanol, with abundances down to $10^{-8}$, before
the onset of the MF and DME recondensations at $70-80$ K.  
Abundance ratios of 100 \% and 10 \% can thus be reached for DME
and MF, respectively when the methanol abundance is still higher than
$10^{-8}$. }

\subsection{Weak and frequent luminosity outbursts}

{ 
Given the lifetime of Class 0 protostars \citep[$\sim 10^5$
yr;][]{Evans2009}, the dynamical timescale of the material in the
envelope inside the centrifugal radius \citep[$10^4 - 10^5$
yr;][]{Visser2009}, and the expected frequency of weak outbursts,
increasing the luminosity by one order of magnitude, of $\sim 5
\times 10^3 - 10^4$ yr \citep{Scholz2013, Vorobyov2015}, it is likely
that cells of material located outside the water snowline undergo
several processes of ice evaporation and condensation.
According to Fig. \ref{nHtemprad}, such luminosity outbursts are
likely strong enough to trigger the evaporation of the whole ice
content into the gas phase up to radii showing a pre-outburst
temperature lower than the condensation temperature of MF and
DME (75 K).    

Figure \ref{X_weakoutbursts} shows the evolution of the absolute
abundances and the abundance ratios during a series of 10 weak
outbursts, in which the temperature is increased from 70 to 125 K
induced by a luminosity increase of one order of magnitude, occurring
every $5 \times 10^3$ yr. 
As for the strong luminosity outburst case, three values of density
were chosen, in order to represent the spread of density expected to
be found at radii where $T = 70$ K in Class 0 and Class I
protostars. 
%
The abundance of MF and DME formed through gas phase chemistry
gradually increases with the outburst number up to one order of
magnitude after 5 outbursts. However, MF and DME are not efficiently
produced during subsequent outbursts because of the gradual
destruction of ammonia during outbursts allowed by the longer total
timescale ($5 \times 10^4$ yr) and its low binding energy that prevents it to
freeze-out.  After five outbursts, the ammonia abundance is already
lower than $10^{-6}$ or 1 \% with respect to water, preventing an
efficient formation of COMs from protonated ions through
proton-transfer reactions (see Fig. \ref{Xch3oh_Xnh3}). 

It can also be seen that the formation efficiency of COMs decreases more
strongly with the density with respect to the strong luminosity outburst
case. The abundances of MF and DME reached after 5 outbursts
decrease by one order of magnitude from $10^{-7}$ to $10^{-8}$ between
$n_{\textrm{H}} = 5 \times 10^6$ and $n_{\textrm{H}} = 5 \times 10^8$
cm$^{-3}$.  
This is due to the very short time spent by CH$_3$OH in the gas phase
at high densities during each weak outburst. 
In total, the CH$_3$OH abundance remains higher than $10^{-6}$ for
more than $10^4$ yr at $n_{\textrm{H}} = 5 \times 10^6$ cm$^{-3}$ but
only for 400 yr at $n_{\textrm{H}} = 5 \times 10^8$ cm$^{-3}$, a shorter
timescale by a factor of 20. In contrast, during one strong luminosity
outburst, CH$_3$OH abundances can remain high for 200 and 1000 yr at
$n_{\textrm{H}} = 5 \times 10^6$ and $5 \times 10^8$ cm$^{-3}$, respectively.
%
However, as it is seen for the strong luminosity outburst case, the
freeze-out timescale decreases with the density to become much shorter
than the outburst timescale for a high density of $5 \times 10^8$ cm$^{-3}$. 
As a consequence, methanol freezes-out efficiently before the onset of
the MF and DME recondensations and their abundance ratios strongly
increase as the methanol decreases, reaching the similar high
abundance ratios as for the strong luminosity outburst case. 

Other parameters can also affect the formation and the freeze-out
efficiencies of COMs and methanol, altering the evolution of their
abundance. 
The outburst timescale or the grain size would influence the
abundances ratios in a similar manner as for the strong luminosity
outburst case. 
The frequency of outbursts is also important for the formation of
daughter COMs through gas phase chemistry. More frequent outbursts
would reinject solid methanol before NH$_3$ is efficiently destroyed by gas
phase chemistry. 
If the ammonia abundance is initially similar to methanol, an increase of
the outburst frequency would tend to increase the formation efficiency
of COMs.  

\begin{figure*}[htp]
\centering 
\includegraphics[height=98mm]{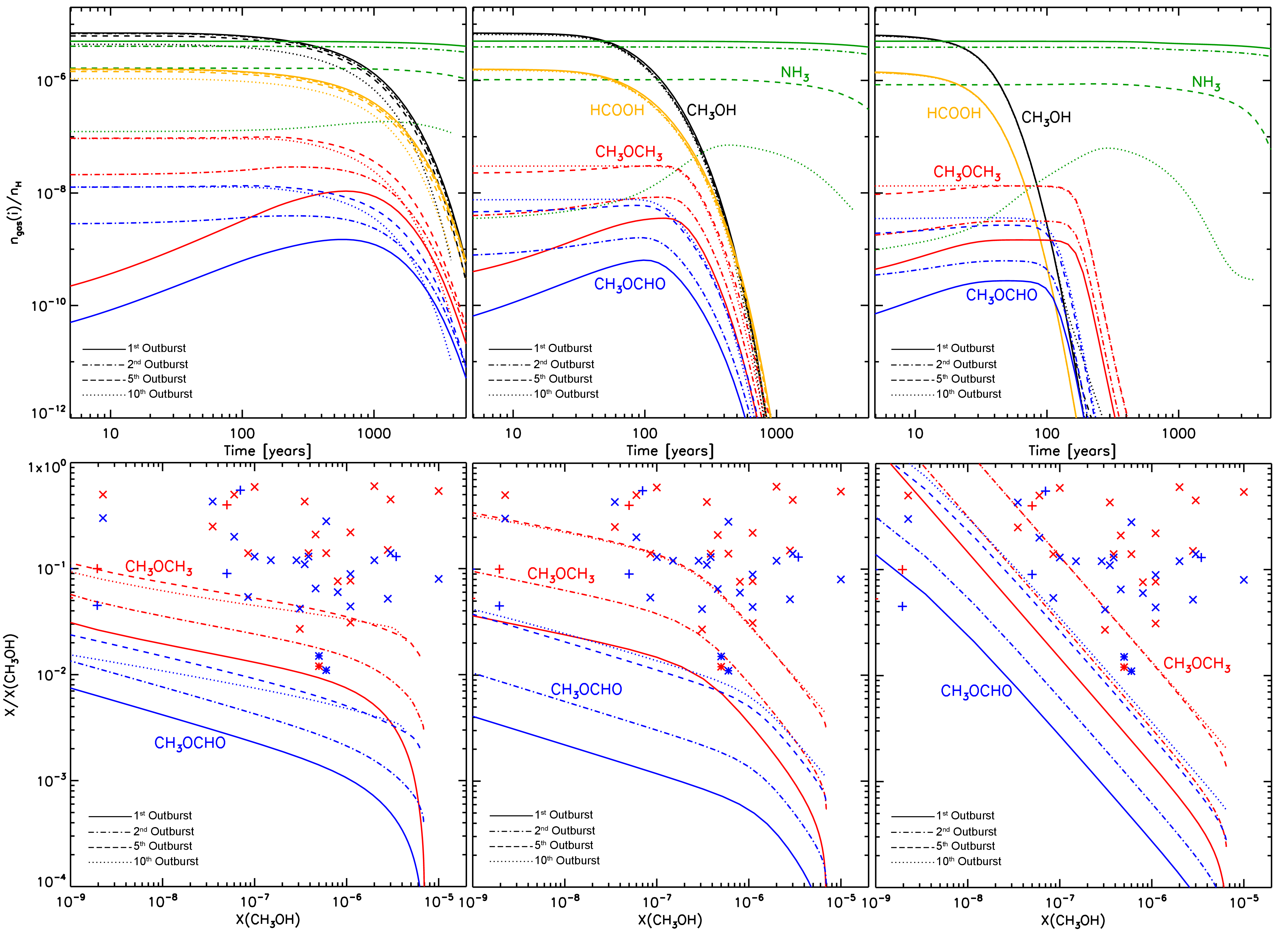}
\caption{
Evolution of the absolute abundances of methanol, ammonia, formic
acid, ethanol, di-methyl ether and methyl formate with time after the
onset of each outburst (top panels) and of the
\ce{CH_3OCH_3}/\ce{CH_3OH} (red) and \ce{CH_3OCHO}/\ce{CH_3OH} (blue)
abundance ratios with the absolute CH$_3$OH abundance (bottom panels)
during a series of 10 weak luminosity outbursts occurring every $5
\times 10^3$ yr assuming $n_{\textrm{H}} = 5 \times 10^6$ cm$^{-3}$
(left), $n_{\textrm{H}} = 5 \times 10^7$ cm$^{-3}$ (center), and
$n_{\textrm{H}} = 5 \times 10^8$ cm$^{-3}$ (right). 
The thickness of the lines increases with the outburst number: the DME
and MF abundances increase with the outburst number while the ammonia
abundance decreases. 
Pluses, stars, and crosses represent the ratios observed toward
low-mass, intermediate-mass, and high-mass protostars, respectively,
summarised in \citet{Taquet2015}.
}
\label{X_weakoutbursts}
\end{figure*}

}

\section{Discussion}

\subsection{Comparison with observed abundance ratios}

{ Gas phase chemistry can produce a large amount of complex
  organics and in particular of DME and MF, the two prototype COMs
  that have been extensively targeted in hot cores, when
  proton-transfer reactions involving ammonia are included . }
In Figure \ref{Xmeth_constant}, observational data obtained toward
more than 40 low-mass to high-mass protostars, are shown together with
the predictions of our static model. 
{ The observational data has been compiled in \citet{Taquet2015},
  and is complemented by the recent detections of ethyl formate toward
  Sgr B2 by \citet{Belloche2009} and of ethyl formate and methyl ethyl
  ether toward Orion KL by \citet{Tercero2015}.}
{ The comparison with observations suggests that gas phase
  chemistry with constant physical conditions is able to explain the
  abundance of DME and MF with respect to methanol toward some of the
  methanol-enriched protostars, with methanol abundances higher than 
$10^{-6}$, that show abundance ratios between 2 and 20 \%. 
However, the abundance ratios higher than 20 \% cannot be reproduced
with our static gas phase model.  

Grain surface chemistry, in which COMs are formed from the UV-induced
radical recombination on warm ($30 \leq T \leq 80$ K) interstellar
grains has been recently proposed to explain the detection of DME, MF and other COMs
around protostars \citep[see][]{Garrod2006, Garrod2008}. 
However, these gas-grain astrochemical models still strongly
underpredict the abundances of DME and MF with respect to methanol by
more than one order of magnitude, the abundance ratios barely
exceeding 1 \% for the two species, suggesting that { other
  chemical processes, such as gas phase chemistry, would play a major
  role for the formation of these} two species  
\citep[see][for a more detailed comparison between observations and
models]{Taquet2015} .     
{ In gas-grain models, complex organic molecules, assumed to be
  mostly formed on interstellar grains, are then evaporated into the
  gas phase when the temperature exceeds $\sim 100$ K and are
  gradually destroyed by gas phase chemistry, through protonation
  followed by dissociative electronic recombination. 
 The incorporation of gas phase proton transfer reactions involving
 NH$_3$ in gas-grain models would delay the destruction of COMs in the gas
 phase, and increase their absolute abundances in hot cores and hot
 corinos, as non-dissociative PT reactions would dominate over the
 dissociative ER reactions.
However, it is unlikely that the abundance ratios of these complex
species with respect to methanol will be increased since COMs and
methanol show similar proton affinities, inducing a similar
chemistry.}  
In particular, the abundance of methyl formate with respect to its
isomer glycolaldehyde CH$_2$OHCHO has been found to be higher than 10
in three low-mass protostars \citep{Jorgensen2012, Coutens2015,
  Taquet2015}, contradicting the gas-grain model predictions of
\citet{Garrod2013}, in which the two molecules are assumed to form
on grains from a similar mechanism, with an abundance ratio
of $\sim 0.1$.  
{ Methyl formate and glycolaldehyde have a similar proton affinity
  of 782 kJ/mol, proton transfer reactions cannot be invoked to
  explain the different abundance ratios. 
As no efficient gas phase formation routes are known for
glycolaldehyde, additional gas phase chemical pathways leading to the
formation of methyl formate naturally explain the high
abundance of methyl formate with respect to its isomer.}

The abundance ratios of ethyl formate and methyl ethyl ether of
$10^{-3}$ observed toward Sgr B2 and Orion KL can also be reproduced
by the gas phase chemistry { depicted in Fig. \ref{chemnet}
  triggered by the evaporation of the   solid methanol, ethanol, and
  formic acid, depending on the assumed physical conditions and
  initial abundances.}
The high proton affinity of ethanol C$_2$H$_5$OH  can explain the lower
abundances of $1-10$ \% relative to methanol observed in the gas phase
of hot cores with respect to the initial abundance of 23 \%  in ices assumed in this
work. However, due to its low proton affinity, the predicted abundance ratio of
HCOOH slightly increases with time and overpredicts the hot core observations
by one to two orders of magnitude. The discrepancy between the model
predictions and the observations could be due to an overprediction
by the models induced by a overestimation of the initial abundance of
HCOOH, or by some missing destruction channels for gaseous
HCOOH. However, it should be noted that the HCOOH abundances derived
from observations in hot cores are only based on the detection of its
more stable {\it trans} conformer.  
Laboratory experiments studying the formation of solid CO$_2$ and
HCOOH through the CO+OH reaction show that the HOCO complex, thought
to act as an intermediate for the formation of both products, can be
formed in its two {\it trans-} and {\it cis-} conformers in similar
quantities \citep{Oba2010, Ioppolo2011}. Moreover, {\it cis}-HCOOH has
recently been detected in a molecular cloud with a similar abundance
than the {\it trans} counterpart (Taquet et al. in prep.) suggesting
that the observed abundance ratios of HCOOH are likely underestimated.} 

{ Due to its high proton affinity, ammonia can abstract a proton from
most of the ions through proton-transfer reactions, altering the
protonation of methanol and the formation of neutral complex organic
molecules from their protonated counterpart. As shown in section 3.2
and Fig. \ref{Xch3oh_Xnh3}, the abundance ratios of COMs highly depend on
the initial abundance of ammonia and reach their maximum at ammonia
abundances of $5-10$ \% with respect to water, when the ammonia
abundance is similar to methanol.}
These abundances are in agreement with the typical icy ammonia abundances
of 0.6-1.4 and 1.3-2.0 relative to solid methanol observed toward
low-mass and high-mass protostars respectively \citep{Oberg2011}.
Gas phase abundances of NH$_3$ and CH$_3$OH should also remain similar as
long as the timescale is not longer than $\sim 10^5$ yr (see Fig. 1). 
We have attempted to derive the NH$_3$/CH$_3$OH abundance ratios
toward 9 high-mass hot cores showing a detection of ammonia and
methanol. 
For all sources, the NH$_3$/CH$_3$OH abundance ratio was derived
following the published NH$_3$ and CH$_3$OH column densities and
scaled according to the size of their emission. We found the following
NH$_3$/CH$_3$OH abundance ratios:
0.38 in G19.61-0.23 \citep{Qin2010}, 
0.37 in G24.78 \citep{Codella1997, Bisschop2007}, 
4.3 in G29.96 \citep{Cesaroni1994, Beuther2007b}, 
6.8 in G31.41+0.31 \citep{Cesaroni1994,   Isokoski2013}, 
0.63 in NGC6334-I-mm1 and 0.32 in NGC6334-I-mm2 \citep{Beuther2007a,
  ZernickelPhD},  
2.5 in NGC7538IRS1 \citep{Bisschop2007, Goddi2015}, 
0.60 in the ``Hot Core'' in Orion KL \citep{Goddi2011, Feng2015}, 
and 0.098 in W33A \citep{Bisschop2007, Lu2014}.
Due to the low number of sources, no trend for the evolution of the
abundance ratio with the methanol abundance can be noticed but we
derived an averaged NH$_3$/CH$_3$OH abundance ratio of 1.8, in good
agreement with the values found in interstellar ices toward high-mass protostars.
The similar abundance of ammonia and methanol found in ices and in the
gas phase suggests that gas phase does not destroy more efficiently
ammonia than methanol in the early stages of star formation. 

{ This work focused on the gas phase chemistry triggered by the
  evaporation of interstellar ices. The chemical composition of ices was
  therefore taken from infrared observations. It is known
  that several species like methanol can show a large variation of
  their abundances with respect to water depending on the source
  \citep{Oberg2011} while the presence of formic acid and ethanol in
  ices inferred from the   band at 7.25 $\mu$m, and their exact
  abundance, is still a matter of debate.   
The gas phase formation of complex organic molecules obviously depends
on the initial abundance of the parent species, the absolute abundance
of dimethyl ether, methyl formate, and other larger species linearly
scales with the initial amount of methanol, formic acid, and ethanol
injected in the gas. 
The variation of methanol abundance does not strongly alter the
abundance ratios of the studied COMs, as long as the abundances of
ethanol, formic acid, and ammonia are scaled to methanol. However,for
a fixed abundance of methanol,  the abundances of methyl formate,
ethyl formate, or methyl ethyl ether tend to vary almost linearly
with the initial abundances of ethanol and formic acid assumed in the ices. 
The high sensitivity and spectral resolution provided by new
generations of infrared telescopes, such as the James Webb Space
Telescope (JWST), together with new laboratory experiments focusing on
the absorption infrared spectra of complex species are therefore
required to confirm the presence of ethanol and formic acid in the
quantities assumed in this work.

We assumed a branching ratio of 100 \% for reaction (\ref{MFreac})
forming {\it trans}-protonated MF following the suggestion by \citet{Cole2012}
(see section 2.2) and for the PT reaction between {\it
  trans}-protonated MF and ammonia, producing  {\it cis}-MF, in the 
absence of quantitative data and based on the energy differences
between {\it cis}- and {\it trans}-MF. 
These branching ratios might be too optimistic and new laboratory work
are needed to confirm or infirm these assumptions. 
Lowering the branching ratio of these reactions to 10 \% decreases the
maximal abundance ratio of methyl formate with respect to methanol by
a factor of 2.5 from 4 \% to 1.5 \% when standard input parameters are
assumed.  
}

\subsection{Impact of luminosity outbursts}


The distribution of bolometric luminosities of embedded protostars
derived by infrared surveys show that most { low-mass} protostars
have relatively low-luminosities of about $1-5 L_{\odot}$
\citep{Evans2009}. 
{
For such low-luminosity sources, the water snow line is located only
10-20 AU away from the central source (or $\sim 0.1$ arcsec at a
typical distance of 200 pc), making the detection of complex organic
molecules in hot cores with current sub-mm facilities very
challenging, even with ALMA. } 
However, in spite of their short timescale of $\sim 100$ years,
luminosity outbursts are able to produce COMs through gas phase
chemistry in significant quantities, with absolute abundances higher
than $10^{-8}$ { in large regions outside the hot core up to 50 to
  200 AU away from the central source, depending on the strength of the
  luminosity outbursts and the structure of the protostellar envelope. 
}

{ 
Due to their low binding energies, the abundances of COMs { formed
  either at the surface of interstellar grains or in the gas phase} relative
to methanol tend to increase after each luminosity outburst as the
methanol decreases. 
In the inner regions of protostellar envelopes, the density varies
between $10^6$ and $10^{10}$ cm$^{-3}$ depending on the distance from
the protostar and the source in consideration. 
Low-density sources would display high absolute abundances of methanol
and COMs long after the luminosity outburst ends due to their slow
freeze-out. At a density of $5 \times 10^6$ cm$^{-3}$, the freeze-out
timescale becomes similar to the timescale between outbursts (1000 - 5000
yr depending on the assumed grain size, see equation \ref{freezeout}), 
methanol and COMs can therefore remain in the gas phase in a large
region of the protostellar envelope during most of the embedded stage.
However, the abundance ratios of COMs would remain limited because
methanol and COMs deplete simultaneously, the abundance ratios after
outbursts do not exceed 5 \%. 
On the other hand, dense protostars that underwent recent outburst
events would likely display a low methanol abundance in the region
just outside of the expected water snow line, due to its fast
freeze-out onto grains, associated with a high abundance ratio of
COMs, induced by their lower binding energy, that could match the
observed abundance ratios. 
According to Fig. \ref{X_outbursts}, the methanol abundance stays
higher than $10^{-10}$ for about 500 yr. Assuming that outburst events
occur every $5 \times 10^3 - 10^4$ yr \citep{Scholz2013, Vorobyov2015}
suggests that an extended emission of COMs could be detected with high
abundance ratios in about $5-10$ \% of dense protostars. 

Other species that show lower binding energies and which also likely undergo
similar behaviours during the recondensation process occurring after
luminosity outbursts can be used to identify chemical clocks for
episodic phenomena. 
By performing SMA observations of the C$^{18}$O emission around a
sample of 16 well-characterised protostars, \citet{Jorgensen2015}
found that half of them show extended C$^{18}$O emission compared to
the C$^{18}$O emission expected from their current luminosities.
This discrepancy can be attributed to previous outburst events
increasing the luminosity by a factor of five or more during the last
$10^4$ yr, and even by a factor of 25 for 25 \% of the observed
sources. 
High-angular resolution observations, using the new generation of
sub-mm interferometers like ALMA, of such sources will be crucial for
testing the gas phase chemistry scenario proposed in this
work. Depending on the density structure of these sources, the
emission of methanol and COMs could eventually be also extended with
respect to the hot core region expected from their current
luminosity. 
Comparison of their emission and their abundance inside and outside
the expected hot core region will help us assessing whether or not
luminosity outbursts can trigger the formation of COMs and alter their
observed abundance ratios.
}

\section{Conclusions}

{
In this work, we have investigated the gas phase formation and
evolution of complex organic molecules (COMs) for constant physical
hot core conditions and during protostellar luminosity outbursts. 
We summarise here the main conclusions of this work:

1) Ion-neutral gas phase chemistry, triggered by the evaporation of
interstellar ices at temperatures higher than 100 K, can efficiently
produce several complex organic molecules. 
{ The incorporation of proton-transfer reactions involving ammonia, in
which its high proton affinity plays a crucial role, results in an efficient
formation and a delayed destruction of complex organic molecules.
The initial abundance of ammonia injected in the gas phase is found to
be the most important parameter for the production of complex organic
molecules.} 
{ These results, in addition to the recent works by \citet{Vasyunin2013}
and \citet{Balucani2015} who proposed new gas phase neutral-neutral
reaction routes, suggest a gas phase origin for several complex
organic molecules.}

2) Comparison with observations suggests that gas phase chemistry
occurring during constant physical conditions can account for the abundances of
  di-methyl ether and methyl formate, the two bright and abundant
  COMs, relative to methanol in almost half of the observed 
protostars without recourse to grain surface chemistry.} In addition,
the abundance ratios of the more complex species ethyl formate and
ethyl methyl ether observed in Orion KL and Sgr B2 can also be
reproduced with our gas phase chemical network. 
{ However, as the gas phase formation of complex organic molecules
  highly depends on the initial abundance of solid species, like HCOOH
  and C$_2$H$_5$OH and the branching ratios of some ion-neutral
  reactions, which are still a matter of debate, more laboratory and
  observational works using new   generations of telescopes are needed
  to confirm these results.} 

3) In spite of their short timescales, { one strong protostellar
  luminosity outburst or a series of five weak outburst events} can produce
complex organic molecules in appreciable amounts through gas phase
chemistry in a large region of protostellar envelopes.   
Di-methyl ether and methyl formate, for example, can be
produced with absolute abundances of about $10^{-8}$ in protostellar
envelope regions with sizes increasing by a factor of 5 to 10,
depending on the strength of the luminosity outburst, with respect to
the pre-outburst hot core. 

4) Because of their lower binding energy that delays their
recondensation, the abundances of di-methyl ether and methyl formate
relative to methanol tend to increase during the cooling occurring
after the outburst, especially when high total densities or low
interstellar grain sizes are assumed.

5) The high abundances of di-methyl ether and methyl formate of $\sim
50$ \% observed toward some of the observed protostars could be
explained  by previous recent luminosity outburst events that trigger
the formation of these molecules in a large region of the envelope
followed by by a delayed recondensation onto grains with respect to
methanol.

~ \\

\begin{acknowledgements}

We thank the anonymous referee for his/her insightful comments that
helped to improve the quality of the manuscript.
V.T. acknowledges the support from the NASA postdoctoral program. 
E.S.W. acknowledges generous financial support from the Swedish
National Space Board. 
S.B.C. was supported by NASA's Origins of Solar Systems Program.

\end{acknowledgements}

\end{document}